\documentclass{svmult}
\pdfoutput=1

\usepackage{psfig}
\usepackage{graphicx}        

\newcommand{\vesc}{\ensuremath{v_{\rm esc}}}

\newcommand{\msun}{\mbox{$M_{\odot}$}}
\newcommand{\lsun}{\mbox{$L_{\odot}$}}
\newcommand{\rsun}{\mbox{$R_{\odot}$}}
\newcommand{\Zsun}{\mbox{$Z_{\odot}$}}
\newcommand{\Teff}{\mbox{$T_{\rm eff}$}}
\newcommand{\vinf}{\mbox{$v_{\infty}$}}

\newcommand{\Cinf}{\mbox{$C_{\infty}$}}
\newcommand{\mdot}{\mbox{$\dot{M}$}}
\newcommand{\ratio}{\mbox{$v_{\infty}$/$v_{\rm esc}$}}
\newcommand{\msunyr}{\mbox{$M_{\odot} {\rm yr}^{-1}$}}

\newcommand{\beq}{\begin{equation}}
\newcommand{\eeq}{\end{equation}}
\newcommand{\beqa}{\begin{eqnarray}}
\newcommand{\eeqa}{\end{eqnarray}}

\newcommand{\eu}{\mbox{${\rm e}$}}
\newcommand{\kms}{\mbox{${\rm km}\,{\rm s}^{-1}$}}

\newcommand{\half}{\mbox{$\frac{1}{2}$}}

\newcommand{\dd}{{\rm d}}

\newcommand{\CIV}{C\,{\sc iv}}

\newcommand{\PV}{P\,{\sc v}}

\newcommand{\Ha} {H$_{\rm \alpha}$}

\newcommand{\logLL}{\mbox{$\log (L/L_{\odot})$}}

\newcommand{\Rstar}{\mbox{$R_{\ast}$}}

\newcommand{\Mstar}{\mbox{$M_{\ast}$}}
\newcommand{\Lstar}{\mbox{$L_{\ast}$}}

\newcommand{\Dmom}{\mbox{$D_{\rm mom}$}}

\newcommand{\vth}{\mbox{$v_{\rm th}$}}

\newcommand{\dvdr}{\mbox{$\dd v/\dd r$}}

\newcommand{\taur}{\mbox{$\tau_{\rm Ross}$}}

\newcommand{\nl}{\mbox{$n_{\rm l}$}}
\newcommand{\taus}{\mbox{$\tau_{\rm Sob}$}}

\newcommand{\qb}{\mbox{$\langle q \rangle$}}
\newcommand{\rhob}{\mbox{$\langle \rho \rangle$}}
\newcommand{\rhobtwo}{\mbox{$\langle \rho^2 \rangle$}}
\newcommand{\gline}{\ensuremath{\mathit{g}_{\rm rad}^{\rm line}}}

\begin{document}

\title*{Mass-loss rates of Very Massive Stars}
\author{Jorick S. Vink}
\institute{Jorick S. Vink\at Armagh Observatory, \email{jsv@arm.ac.uk}}

\maketitle

\abstract{We discuss the basic physics of hot-star winds and we provide 
mass-loss rates for (very) massive stars. 
Whilst the emphasis is on theoretical concepts and line-force modelling, we 
also discuss the current state of observations and empirical modelling, and address the issue of wind clumping.} 

\section{Introduction}
\label{sec:intro}

Mass loss via stellar winds already plays a well-documented role 
in the evolution of canonical 20-60\,\msun\ O stars, because of the removal 
of {\it mass} from the outer layers, as well as  
the removal of {\it angular momentum}. 
However, nowhere is mass loss more dominant than for the most massive stars. 
As very massive stars (VMS) evolve structurally 
close to chemically homogeneously, the detailed mixing processes 
due to rotation and magnetic fields are less relevant than for canonical massive stars. 
Instead, VMS evolution 
is determined by mass loss (Yungelson et al. 2008; Yusof et al. 2013; K\"ohler et al. 2014).
However, there is uncertainty regarding the quantitative 
mass-loss rates, partly because of uncertain physics in close proximity to the 
Eddington ($\Gamma$) limit, and partly because 
O-star winds are inhomogeneous and clumpy, implying that empirical mass-loss rates are overestimated 
if one does not properly take clumping effects into account in the analysis. 

In this mass-loss chapter, we start off in Sect.\,2 with the mass-loss theory 
of canonical 20-60\,\msun\ O-star winds, which are optically thin, and 
where the traditional CAK theory due to Castor, Abbott \& Klein (1975)
is applicable. 
For VMS, the role of radiation pressure over gas pressure
is even more important than for normal massive stars, and as VMS are in closer proximity to the $\Gamma$ limit, 
at some point their winds are expected to become optically thick. 

In Sect.\,3, we discuss the optically thick wind theory  
for classical Wolf-Rayet (WR) stars with very strong emission lines and dense winds. 
Once we have reached a basic understanding of both optically thin 
and optically thick winds\footnote{Note however that these winds 
are also driven by a myriad of lines, forming a ``pseudo'' continuum of lines.}, we 
discuss the transition from O to WR star winds in the context of VMS in Sect.\,4.
VMS are associated with WR stars of the WNh subtype. WNh implies 
the presence of both hydrogen (H) and nitrogen (N) at the surface. 
The latter is thought to have originated in the 
CNO-cycle, and reaching the surface through mass loss and (rotational) mixing. 
WNh stars are thought to be 
core H-burning (see Martins' Chapter 2) and can thus be considered ``O-stars on steroids''. 
The reason they have a WR type spectrum is due to their strong winds, because of the
proximity to the Eddington limit.

Another group of objects that may be relevant for VMS evolution are 
the Luminous Blue Variables (LBVs). Already in quiescence these objects reside in dangerous proximity 
to the Eddington limit, where they are subjected to outbursts and 
mass ejections. A discussion 
of both ``quiet'' and super-Eddington winds relevant to both the 
characteristic  ``moderate'' S\,Dor variations and the ``giant'' outbursts, such 
as displayed by Eta Car in 1840, as well as the theory of super-Eddington winds
are thus discussed in Sect. 6.

After this theoretical overview of homogeneous stellar winds, 
we consider clumped winds. To this purpose, we first discuss the 
diagnostics of smooth winds (Sect.\,7) before turning to clumped winds in Sect.\,8. 
We finish Sect.\,8 with potential theories that may cause 
wind clumping, as well as some possibilities to quantify the number of clumps, before 
we summarise in Sect.\,9. 

For the 2D effects of rotation on stellar winds, we refer to the review by 
Puls et al. (2008) and for more recent calculations to 
M\"uller \& Vink (2014), which also includes 
a discussion of the diagnostics of axi-symmetric outflows.

\section{O stars with optically thin winds}
\label{sec:ldw}

As each photon carries a momentum, $P = h \nu/c$, it was thought as early as 
the 1920s (e.g. Milne 1926) that radiative acceleration on 
spectral lines might selectively ``eject'' metal ions (such as iron, Fe) 
from stellar photospheres.
However, it was not until the arrival of 
ultraviolet (UV) observations in the late 1960s 
that the theory of radiative line driving became the established 
theory describing the stationary outflows from massive OB stars. 
Lucy \& Solomon (1970) and CAK showed that in case 
the momentum imparted on metal ions was shared through Coulomb 
interactions with the more abundant H and helium (He) 
species\footnote{Note that for every Fe atom there are as many as 
2500 H atoms (for a solar abundance pattern; see Anders \& Grevesse 1989).} 
in the atmospheric 
plasma, this would result in a substantial rate of mass loss \mdot, 
affecting the evolution of massive stars significantly (Conti 1976; Langer et al. 1994; 
Meynet \& Maeder 2003; Eldrige \& Vink 2006; 
Limongi \& Chieffi 2006; Belkus et al. 2007; Brott et al. 2011; Hirschi's Chapter 6; and Woosley \& Heger's Chapter 7).

\subsection{Stellar wind equations}
\label{sec:standardmodel}

The basic idea of momentum transfer by line-scattering is that 
absorbed photons originate from a preferred direction, whereas
the subsequent re-emission is averaged to be (more or less) isotropic. 
This change in direction angle $\theta$ leads to a {\it radial} 
transfer of momentum, $\Delta P = h/c (\nu_{\rm in} \cos \theta_{\rm in} -
\nu_{\rm out} \cos \theta_{\rm out})$ -- comprising the key to the 
momentum transfer with its associated line acceleration 
$g_{\rm rad}^{\rm line}$. 
The mass-loss rate through a spherical shell with radius $r$ 
that surrounds the star is conserved, as may be noted from
the equation of mass continuity
\begin{equation}  
\label{eq_const}
\dot{M} =  4\,\pi\, r^{2}\, \rho\,(r)\,  v\,(r).
\end{equation}
The equation of motion is given by:
\begin{equation}
v \frac{\dd v}{\dd r}~=~- \frac{G M}{r^2}~-~\frac{1}{\rho}\frac{\dd p}{\dd
r}~+~g_{\rm rad},
\label{eq_mom}
\end{equation}
with inwards directed gravitational acceleration $g_{\rm grav} = GM/r^2$ 
and an outwards directed
gas pressure ($p$) term and total (continuum and line) 
radiative acceleration ($g_{\rm rad}$). 
The wind initiation condition is that the total
radiative acceleration, $g_{\rm rad}$ = $g^{\rm line}_{\rm rad}$ $+$ $g^{\rm
cont}_{\rm rad}$ exceeds gravity beyond a certain point. 
With the equation of state, $p = a^{2}\,\rho$,
where $a$ is the isothermal sound speed, the
equation becomes:
\begin{equation}
\bigl(1-\frac{a^2}{v^2}\bigr)~v \frac{\dd v}{\dd r}~=~
\frac{2a^2}{r}~-~\frac{\dd a^2}{\dd r}~-~\frac{G M}{r^2}~~+~g_{\rm rad}.
\label{eq_motion}
\end{equation}

\noindent
The prime challenge lies in accurately computing $g_{\rm rad}$. 
For free electrons this concerns the Thomson opacity, $\sigma_{\rm e} = s_{\rm e} \rho$ ($s_{\rm e}$
proportional to cross section) and the flux:
\begin{equation}
g^{\rm Th}_{\rm rad}~=~\frac{1}{c \rho} \frac{\sigma_{\rm e} L}{4 \pi r^2}~=~
g_{\rm grav}\,\Gamma,
\label{eq_elec}
\end{equation}
with the Eddington parameter $\Gamma$ representing the radiative acceleration over gravity, 
given by:

\begin{equation}
\label{eq_gamma}
  \Gamma = \frac{\kappa L}{4\pi c GM}.
\end{equation}
Spectral lines provide the dominant contribution to the overall radiative acceleration. 
The reason is that line scattering is intrinsically much stronger than electron
scattering because of the resonant nature of bound-bound transitions
(Gayley 1995), and although photons and matter are only allowed
to interact at specific frequencies, they can be made to resonate over a
wide range of stellar wind radii via the Doppler effect (see Owocki's Chapter 5). 

For a single line at frequency $\nu$, with line optical depth $\tau$,
the line acceleration can be approximated by local quantities (Sobolev 1960). This
approximation is valid as long as opacity, source function, 
and the velocity gradient ($dv/dr$) do not change significantly over a velocity range
$\Delta v = \vth$, corresponding to a {\it spatial} region 
$\Delta r \approx \vth/(\dvdr)$, i.e. the Sobolev length.
In the Sobolev approximation, the line acceleration becomes:
\begin{equation}
g^{\rm line}_{\rm rad, i}~=~\frac{L_\nu \nu}{4 \pi r^2 c^2}~(\frac{\dd v}{\dd
r})~ \frac{1}{\rho}~(1~-~e^{-\tau}),
\label{eq_gline}
\end{equation}
with $L_\nu$ the luminosity at the line frequency, and with 
\beq
\tau = \bar \kappa \lambda /(\dvdr), 
\label{tausob}
\eeq
where $\bar \kappa$ represents the frequency
integrated line-opacity and $\lambda$ is the wavelength of the transition. 
For optically thin lines ($\tau < 1$) the line acceleration 
has the same $1/r^2$ dependence as electron scattering (Eq.~\ref{eq_elec}), whereas
for optically thick lines ($\tau > 1$) it depends on 
the velocity gradient (\dvdr), which is the root cause for the peculiar nature of line driving.

\subsection{CAK solution}

The next step is to sum the line acceleration over all lines. In the CAK 
theory this is achieved 
through the line-strength distribution function that describes the
statistical dependence of the number of lines on frequency position and
line-strength (e.g. Puls et al. 2000). 
Combining the radiative line acceleration (Eq.~\ref{eq_gline}) with the
distribution of lines, the {\it total} line acceleration can be
calculated by integration. It can be expressed in terms of the
Thomson acceleration (Eq.~\ref{eq_elec}) multiplied by the 
famous force-multiplier $M(t)$,
\begin{equation}
M(t)~=~\frac{g^{\rm line}_{\rm rad}}{g^{\rm TH}_{\rm rad}}~=~k~t^{-\alpha}
\propto \bigl(\frac{\dvdr}{\rho}\bigr)^\alpha  ,
\label{eq_cak}
\end{equation}
where $k$ and $\alpha$ are the so-called force multiplier parameters.

For the complete distribution of lines, the radiative 
acceleration depends on ($\dvdr$) through the power of $\alpha$.
CAK postulated that this term has a similar
meaning as the velocity gradient entering the inertial term on the left hand
side of
Eq.~\ref{eq_motion}. Assuming this is the 
case, the equation of motion becomes non-linear, and can be solved
through a critical \label{critpoint} point that sets the mass-loss 
rate \mdot:
\beq
\dot{M} \propto (kL)^{1/\alpha}~(M(1-\Gamma))^{1-1/\alpha}.
\label{eq_mdot}
\eeq
And with velocity:
\begin{equation}
v(r) = \vinf (1~-~R/r)^{\beta}
\label{eq_vr}
\end{equation}
\begin{equation}
\vinf = \Cinf \bigl(\frac {2 G M (1-\Gamma)}{\Rstar} \bigr)^{\half} = \Cinf \vesc ,
\label{eq_vinf}
\end{equation}
where \Cinf\,$\approx$\,2.6 for O stars, and \vesc\ is the 
photospheric escape velocity corrected for Thomson electron 
acceleration. $\beta$ is exactly 0.5 for a point source, and in the range 
$\beta\,\approx\,0.8-1$ for more realistic (finite sized) objects 
(Pauldrach et al. 1986; M\"uller \& Vink 2008). 
For O stars, $\alpha \simeq0.6$ and $k$ is of the order of 0.1. 

Using these relations, one can construct the modified wind momentum rate,
$\Dmom\,=\,\mdot\,\vinf\,(\Rstar/\rsun)^{1/2}$. Given that \vinf\ scales
with the escape velocity (Eq.~\ref{eq_vinf}), \Dmom\ scales with luminosity
and effective line number only, and as long as $\alpha$ $\simeq$ $2/3$, the effective 
mass $M(1-\Gamma)$ conveniently cancels from the product $\mdot \, \vinf$, resulting in: 
\begin{equation}
\log \Dmom~\approx~x~\logLL + D ,
\label{wlr}
\end{equation}
(with slope $x$ and offset $D$, depending on the flux-weighted number of
driving lines), the ``wind momentum
luminosity relationship (WLR)'' 
(Kudritzki et al. 1995; Puls et al. 1996; Vink et al 2000). 
The relationship played an instrumental role in determining the
empirical mass-loss metallicity ($Z$) dependence for O stars in 
the Local Group (Mokiem et al. 2007), and observed and predicted WLRs 
can be compared to test the validity of the theory, and to highlight potential 
shortcomings, e.g. concerning wind clumping. One should also realise that 
$\dot{M}$ is not only a function of $L$ but also parameters like
$T_{\rm eff}$. One should properly account for this multivariate behaviour of $\dot{M}$ 
when one attempts to compare observations to theory, and when one wishes to 
properly assess the effects of stellar wind mass loss in stellar evolution modelling.

We note that all CAK-type relations are only valid 
for spatially constant force multiplier parameters, $k$ and $\alpha$, 
which is not the case in more realistic models 
(Vink 2000; Kudritzki 2002; Muijres et al. 2012a).
Other assumptions involve the adoption of a core-halo structure, and the 
neglect of multi-line effects. 

\subsection{Predictions using a Monte Carlo radiative transfer approach}
\label{sec:statmodels}

\begin{figure}
\begin{center}
   {\includegraphics[width=8cm]{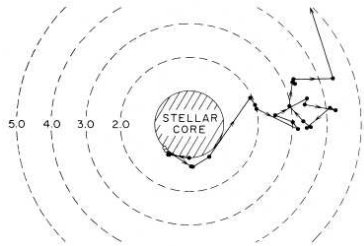}}
\end{center}
\caption{Cartoon explaining the Monte Carlo method: photon path histories are 
tracked on their outwards journey. From Abbott \& Lucy (1985).}
\label{fig_al85}
\end{figure}

An alternative approach to CAK involves the Monte Carlo method developed by 
Abbott \& Lucy (1985). Here photon-scattering histories are 
tracked on their journey outwards. At each interaction, momentum and energy 
are transferred from the photons to the ions (see Fig.\,\ref{fig_al85}). 
One of the major advantages of the Monte Carlo method is that it easily
allows for multi-line scattering, which becomes important in denser winds. 
Prior to the year 2000, theoretical mass-loss rates fell short of the observed rates
for dense O star and WR winds, whilst for weak winds the oft-used single
line approach overestimated mass-loss rates. 
The crucial point is that multiply scattered photons add radially outward
momentum to the wind, and the momentum may exceed the 
single-scattering limit, i.e., $\eta = \mdot \vinf/(L/c)$ can become 
larger than unity. The overall $\dot{M}$ 
can be obtained from global energy conservation:
\begin{equation}
\frac{1}{2} \mdot (\vinf^2 + \vesc^2) = \Delta L,
\label{energy}
\end{equation}
where $\Delta L$ is the total energy transferred per
second from the radiation to the outflowing particles. 

\begin{figure}
\begin{center}
   {\includegraphics[width=8cm]{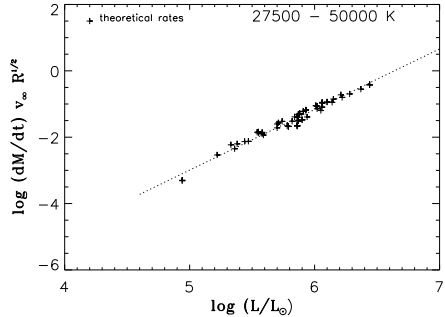}}
\end{center}
\caption{Predicted WLR for O stars hotter than 27~kK for 
a range of $(L,M)$-combinations in the upper HR diagram. From Vink et al. (2000).}
\label{wlr_vink}
\end{figure}

Vink et al. (2000, 2001) used the Monte Carlo method to derive 
a mass-loss recipe, where for objects hotter than the 
so-called bi-stability jump at $\simeq 25~000$ K, the rates roughly 
scale as:
\begin{equation}
\mdot~\propto~L^{2.2}~M^{-1.3}~\Teff~(\ratio)^{-1.3}.
\label{eq_formula}
\end{equation}
The success of the Monte Carlo method is highlighted through the comparison
of observed and predicted mass-loss rates in Vink (2006). 
Figures\,1 and\,4 of that review display the level of agreement 
between modified CAK models and observations on the one hand, and the 
Vink et al. (2000) predictions on the other hand. 
Despite remaining uncertainties due to an unknown amount of wind clumping, by 
properly including multiple scatterings, the results were shown to be  
equally successful for relatively 
weak (with \mdot\ $\sim$ $10^{-7}$ \msunyr) as dense O-star 
winds (with \mdot\ $\sim$ $10^{-5}$ \msunyr). 
The predictions can also be expressed via the WLR. 
For O-stars hotter than 27\,500 K, the relation is shown in Fig.~\ref{wlr_vink} 
and given by Eq.~(\ref{wlr}) with a slope $x$ $=$ 1.83.  

Traditionally, the prime drawback of the Monte Carlo approach was 
the usage of a pre-determined $\vinf$ (guided 
by accurate empirical values) but this assumption can be 
dropped, as discussed in the following. 

\subsection{Line acceleration formalism $g(r)$ for Monte Carlo use}

In solving the equation of motion self-consistently without relying on 
any free parameters, M\"uller \& Vink (2008) determined the velocity field 
through the use of a parameterised description of the line acceleration 
that only depends on radius (rather than explicitly on the velocity gradient dv/dr as in CAK theory.) 
The line acceleration was obtained from Monte Carlo radiative transfer calculations. 
As this acceleration is determined in a statistical way, it shows scatter, and 
given the delicate nature of the equation of motion it should 
be represented by an appropriate analytic fit function. 
M\"uller \& Vink (2008) motivated:
\beq
\label{eq:grad}
\gline = \left\{ \begin{array}{rl} 
       0 & \hspace{6mm} \textrm{if  } \hspace{2mm} r < r_{\circ} \\
       \mathit{g}_{\circ} \,(1-r_{\circ}/r)^{\gamma} / r^{2} & 
       \hspace{6mm} \textrm{if  } \hspace{2mm} r \geq r_{\circ}, \\
	              \end{array} \right.		
\eeq
where $\mathit{g}_{\circ}$, $r_{\circ}$, and $\gamma$ are fit parameters to the Monte Carlo 
line acceleration. 
M\"uller \& Vink (2008) derived an analytic solution of the velocity law in the 
outer wind, which was compared to the standard CAK $\beta$-law and subsequently 
used to derive \vinf\ and the most representative $\beta$ value. 

Equation \ref{eq_motion} is a critical point equation, where the left- and right-hand 
side vanish at the point $v(r_s)=a_{\circ}$, i.e. where $r_s$ is the sonic-point radius. 
M\"uller \& Vink (2008) showed that for the isothermal case and 
a line acceleration as described in Eq.~\ref{eq:grad}, analytic expressions for 
all types of solutions of Eq.~\ref{eq_motion} can be constructed by means of the 
Lambert W function. 
A useful approximate wind solution for the velocity law 
can be constructed if the gas pressure related terms $2a^{2}/r$ and
$a/v$ are neglected. After some manipulation one obtains 
the approximate velocity law:
\beq
\label{eq:vlawapprox}
v(r) = \sqrt{ \frac{R_* v_{\rm esc}^2}{r} +  \frac{2}{r_{\circ}}\frac{\mathit{g}_{\circ}}{\left( 1+\gamma \right)} \left(1-\frac{r_{\circ}}{r} \right)^{\gamma + 1} + C},
\eeq
where $C$ is an integration constant. From this equation the terminal wind 
velocity can be derived if the integration constant $C$ can be determined, which  
can be done by assuming that at radius $r_{\circ}$ the velocity approaches zero, resulting in:
\beq
\label{eq:integrationc}
C = - \frac{R_* \vesc^2}{r_{\circ}}.
\eeq
In the limit $r \rightarrow \infty$:
\beq
\label{eq:vinf}
\vinf = \sqrt{\frac{2}{r_{\circ}} {\frac{\mathit{g}_{\circ}}{(1+\gamma)} - \frac{R_* \vesc^2}{2}}}.
\eeq

\noindent
The terminal velocity $\vinf$ can also be determined from the equation of motion. 
At the critical point, the left-hand and right-hand side of Eq.~\ref{eq_motion} both equal zero. Introducing
\vinf\ in relation to $\mathit{g}_{\circ}$ as expressed in Eq.~\ref{eq:vinf}, one obtains
\beq
\label{eq:vinfnew}
v_{\infty,{\rm new}} = \sqrt{\frac{2}{r_{\circ}} \left[ \left( \frac{r_s}{r_s-r_{\circ}} \right)^{\gamma} 
                 \frac{r_s}{(1+\gamma)} \left( \frac{\vesc}{2} - 2 r_s \right) - \vesc^2 \right]}.
\eeq
A direct comparison to the $\beta$-law can be made for the 
supersonic regime of the wind, resulting in 
\beq
\label{eq:beta}
\beta = \frac{1+\gamma}{2}.
\eeq

\noindent
The procedure to obtain the best-$\beta$ solution is that
in each iteration step of the 
Monte Carlo simulation the values of $\mathit{g}_{\circ}$, $r_{\circ}$, and $\gamma$ are 
determined by fitting the output line acceleration. 
Using these values and the radius of
the sonic point, Eqs.~\ref{eq:vinf},~\ref{eq:vinfnew} and~\ref{eq:beta} are used to determine
$\vinf$ and $\beta$. \vinf\ derived from Eq.~\ref{eq:vinfnew}, the predicted mass-loss rate, and 
the expression derived for $\beta$ serve as input for the next 
model, with iterations continuing until convergence is achieved.

Muijres et al. (2012a) tested the M\"uller \& Vink (2008) 
wind solutions through explicit numerical integrations of the 
fluid equation, also accounting for a temperature stratification, obtaining 
results that were in excellent agreement with the M\"uller \& Vink solutions. 
These solutions were extended to 2D in M\"uller \& Vink (2014).

\section{Wolf-Rayet stars with optically thick winds}
\label{sec_thick}

\begin{figure}
\begin{center}
   {\includegraphics[width=8cm]{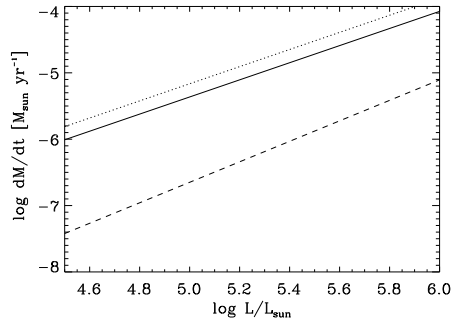}}
\end{center}
\caption{Comparison of mass-loss rates from WR and Galactic O supergiants (from 
Puls et al. 2008). 
Solid and dotted lines represent mean relations for H-poor WN (solid) and WC
stars (dotted) from Nugis \& Lamers (2000).
The dashed line corresponds to Galactic O supergiants -- taken
from the Mokiem et al. (2007) WLR.}
\label{mdot_wr}
\end{figure}

\subsection{Wolf-Rayet (WR) stars}

WR stars can be divided into nitrogen-rich WN stars and carbon/oxygen 
rich WC/WO stars. The principal difference between the two
subtypes is believed to be that the N-enrichment in WN stars is a 
by-product of H-burning, whereas the C/O in WC/WO stars is due to the arrival of 
He-burning products at the surface, showing strong emission lines of He, C and O. 

The WR classification is purely
spectroscopic, signalling the presence of strong and broad
emission lines. 
Such
spectra can originate in evolved stars, or alternatively from objects that 
formed with high initial masses and luminosities, the VMS. 
This latter group of WR stars may
thus include objects still in their core H-burning phase of evolution: WNh stars.

Stellar radii determined from sophisticated non-LTE models are 
a factor of several ($\sim$3) larger
than those predicted for the He-main sequence by stellar evolution modelling.
In other words, there is a radius problem, and a potential 
solution might involve the inflation of a clumped outer envelope (Gr\"afener et al. 2012; 
See Chapter\,5 for more details). 

\subsection{WR wind theory}  

WR stars have strong winds with large mass-loss rates, typically a factor of 
10 larger than O-star winds with the same luminosity (see Fig.\,\ref{mdot_wr}), 
and they are not easily explained by the optically thin 
line-driven wind theory by CAK.
The observed wind efficiency $\eta$ values are
typically in the range of 1-5, i.e. well above the single-scattering limit.
So, {\em if} WR-type winds are driven by radiation, photons must be scattered more 
than once.
As the ionisation
equilibrium decreases outwards, photons can
interact with lines from a variety of different ions on their way out, whilst gaps 
between lines become ``filled in'' (see Lucy \& Abbott 1993; Schaerer \& Schmutz 1994; 
Springmann 1994; Gayley et al. 1995). 

The initiation of the mass loss relies on
the condition that the winds are already optically {\it thick} 
at the sonic point and that the photospheric line acceleration due
to the high opacity ``iron peak'' may overcome gravity, thus driving a wind (Nugis \& Lamers 2002). 

The crucial point in such a critical-point analysis for optically thick winds  
is that due to their large mass-loss rates, the atmospheres become so 
extended and the sonic point of the wind is already reached 
at large flux-mean optical depth $\tau_s$, which implies 
that the radiation can be treated in the diffusion approximation. 
The equation for the radiative acceleration can then be approximated to:

\begin{equation}
 \label{eq:arad}
  g_{\rm rad} = \frac{1}{c} \int \kappa_\nu F_\nu {\rm d}\nu
  = \kappa_{\rm Ross}\frac{L_\star}{4\pi r^2 c},
\end{equation}
where $\kappa_{\rm Ross}$ is the Rosseland mean opacity which can be
taken from for instance the OPAL opacity tables (Iglesias \& Rogers 1996). 
As $g_{\rm rad}$ does not depend on ($\frac{{\rm d}v}{{\rm d}r}$) 
Eq.\,(\ref{eq_motion}) has a
critical point at the sonic point $r_{\rm s}$ where $v=a$. 

A finite value of
($\frac{{\rm d}v}{{\rm d}r}$) can only be obtained if the right hand side of
Eq.\,(\ref{eq_motion}) is zero at this point.

\begin{equation}
\label{eq:crit}
  0 = -\frac{GM}{r_{\rm s}^2} 
  + \frac{2 a^2}{r_{\rm s}} - \frac{{\rm d}a^2}{{\rm d}r_{\rm s}} + g_{\rm rad}.
\end{equation}
For reasonable wind parameters the second and third term on 
the right-hand side of
Eq.\,(\ref{eq:crit}) become zero such that
\begin{equation}
\label{eq:ledd}
  \frac{GM}{r_{\rm s}^2} \simeq g_{\rm rad}(r_{\rm s})
  \equiv \kappa_{\rm crit}\frac{L_\star}{4\pi r_{\rm s}^2 c}.
\end{equation}
The Eddington limit with respect to the Rosseland mean opacity is thus 
crossed at the sonic point, and $\kappa_{\rm crit}$ for the
Rosseland mean opacity can be computed for stellar parameters in terms 
of the ($L/M$) ratio.

\begin{figure}
\begin{center}
   {\includegraphics[width=8cm]{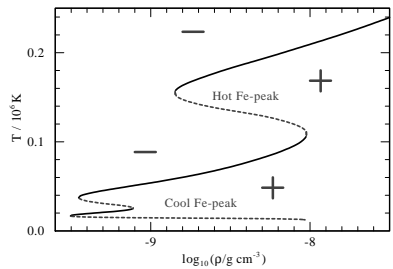}}
\end{center}
\caption{Solution of Eq.\,(\ref{eq:ledd}) in the $\rho$-$T$ plane. The sonic-point
  conditions for an optically thick wind, i.e.\ $\kappa_{\rm Ross}=\kappa_{\rm
    crit}$ with {\em outward increasing} $\kappa_{\rm Ross}$, are fulfilled
  at the solid parts of the curve around 70\,kK, and above 160\,kK. The
  Rosseland opacities are taken from the OPAL opacity tables.
  From Gr\"afener \& Hamann.}
  \label{fig1}
\end{figure}

In Fig.\,\ref{fig1} the solution of Eq.\,(\ref{eq:ledd}) is plotted. This 
figure shows the relation between density and temperature with $\kappa_{\rm Ross}(\rho,
T)= \kappa_{\rm crit}$, for a typical WC star. 
Below the sonic point, $r_{\rm s}$, the radiative acceleration must be sub-Eddington, and 
$\kappa_{\rm Ross}$ thus needs to increase
outward with decreasing density. 
Figure\,\ref{fig1} shows how this condition is fulfilled at the hot edges of 
two Fe opacity peaks, one ``cool'' one at $\sim$ 70\,kK and a ``hot'' one 
above 160\,kK.  The resulting
mass-loss rates on these parts of the curve are given by $\dot{M}= 4\pi
R_\star^2\rho\,a$.
To determine the actual density and temperature at the sonic point, 
Nugis \& Lamers (2002) utilised the approximate relation between temperature 
and optical depth due to Lucy (1971) (see also Gr\"afener \& Vink 2013):

\begin{equation}
\label{eq:lucy}
 T_{\rm S}^4(r) = \frac{3}{4}T_{\rm eff}^4 \left( \tau_{\rm S}(r)+\frac{4}{3}W(r)\right),
\end{equation}
with the modified optical depth $\tau_{\rm S}$ and the dilution factor $W$, which is
close to unity.  
$\tau_{\rm S}$ is obtained from the assumption that the
outer wind is driven by radiation, and by combining Eqs. (34) and (35) of Nugis 
\& Lamers (2002)
for the optical depth and the temperature stratification, the 
resulting mass-loss rate for optically-thick winds is:

\begin{displaymath}
\mdot = C \frac{a T_{\rm s}^{4} R_{\rm S}^{3}}{M}.
\end{displaymath}
Nugis \& Lamers (2002) found that the observed WR
mass-loss rates are in agreement with this optically-thick wind assumption,
and with a {\it bifurcation} of two sonic-point temperature regimes: 
a ``cool'' regime corresponding to late-type WN (WNL) 
stars, and a hot regime for early-type WR (WC and WN) stars.  

\subsection{Hydrodynamic optically thick wind models}

Gr\"afener \& Hamann (2005) included the OPAL Fe-peak opacities of 
the ions Fe\,{\sc ix--xvii} in more sophisticated models that treat 
the full set of non-LTE population numbers in combination with 
the radiation field in the co-moving frame (CMF). 
Combining these models with the equations of hydrodynamics, Gr\"afener \& Hamann
obtained a self-consistent model for the WC5 star WR\,111. 
The resulting wind acceleration and Fe-ionisation structure are depicted
in Fig.\,\ref{acc}.
$g_{\rm rad}$ was obtained from an integration of the
product of opacity and flux over frequency (see Eq.\,{\ref{eq:arad}). 
Wind clumping was treated in the optically thin (``micro'') clumping 
approach (see Sect.\,8.1). 
With a mass-loss rate of $\dot{M}=10^{-5.14}\,M_\odot/{\rm yr}$
and terminal wind velocity of $v_\infty=2010\,{\rm km}/{\rm s}$, the
observed spectrum was also reproduced, although 
the electron scattering wings highlighted that the assumed 
clumping factor of $D=50$ was rather (too) large 
given that WC\,stars generally seem to have clumping factors 
of the order of $D=10$, as determined from electron scattering wings 
(Hillier 1991; Hamann \& Koesterke 1998). 
The models might
therefore underestimate the mass loss rate by a factor of $\sqrt{5}$.
This is likely due to the omission 
of opacities of intermediate-mass elements, such as 
Cl, Ne, Ar, S, and P, which according to 
Monte Carlo models may account for up to half of 
the total line acceleration in the outer wind (Vink et al. 1999).

\begin{figure}
\begin{center}
   {\includegraphics[width=8cm]{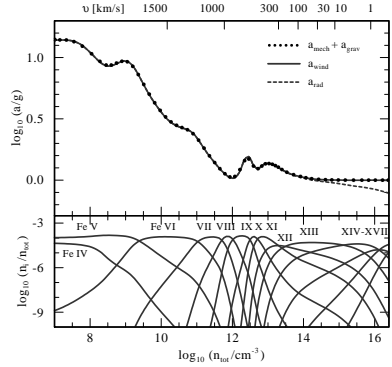}}
\end{center}
\caption{
  Top panel: the radiative acceleration of the Gr\"afener \& Hamann (2005) 
  WC5 star model WR\,111 (expressed in units of
  the local gravity). The wind acceleration $g_\mathrm{wind}$ due to
  radiation and gas pressure balances the mechanical and 
  gravitational acceleration $g_\mathrm{mech} + g_\mathrm{grav}$. Bottom panel:
  the Fe-ionisation structure.
  \label{acc}}
\end{figure}

\section{VMS and the transition between optically thin and thick winds}
\label{sec_thick}

There are many uncertainties in the quantitative 
mass-loss rates of both VMS as well as canonical 20-60\,\msun\ massive stars. 
One reason is related to the role of wind clumping, which will be discussed later, but 
there are also uncertainties related to modelling techniques. 
Nevertheless, arguably the most pressing 
uncertainty is actually still qualitative! 
Do VMS winds become optically thick in Nature? 
\footnote{Note that Pauldrach et al. 2012 argue that VMS winds 
remain optically thin.} And if so, would this lead to an accelerated 
increase of $\dot{M}$? And if so, at what point does the 
transition occur?

\subsection{Analytic derivation of transition mass-loss rate}

As hydrostatic equilibrium is a good approximation 
for the subsonic part of the wind the terms on the right-hand side
of Eq.\,\ref{eq_mom} cancel each other. 
In the supersonic portion of the wind the gas pressure 
gradient becomes small, and through multiplying Eq.\,\ref{eq_mom} by $4 \pi r^2$, it reads:

\begin{equation}
4 \pi \rho r^2 v dv = 4 \pi r^2 (g_{\rm rad} - g) dr.
\label{eq_LC99}
\end{equation}
Employing the mass-continuity equation, one obtains

\begin{equation}
\dot{M} dv = 4 \pi G M (\Gamma(r) - 1)\rho dr. 
\end{equation}
Where $\Gamma(r)$ the Eddington
factor with respect to the total flux-mean opacity $\kappa_{\rm F}$: 
$\Gamma(r) = \frac{\kappa_{\rm F} L}{4\pi c GM}$. Using the wind optical depth 
$ \tau = \int_{r_s}^\infty \kappa_{\rm F}\rho\, {\rm d}r$, one obtains

\begin{equation}
\frac{\dot{M}}{L/c} dv = \kappa_{\rm F} \rho \frac{\Gamma - 1}{\Gamma} dr = \frac{\Gamma-1}{\Gamma} d\tau.
\label{eq_gaga}
\end{equation}
Assuming hydrostatic equilibrium below the sonic point, in integral form this becomes:

\begin{equation}
\int_0^{v_{\infty}} \frac{\dot{M}}{L/c} dv = \frac{\dot{M} v_{\infty}}{L/c} = \int_{r_{\rm S}}^{\infty} \frac{\Gamma - 1}{\Gamma} d\tau 
\simeq \tau_{\rm S}.
\label{eq_gaga}
\end{equation}
Where it is assumed that $\Gamma$ is
significantly larger than one in the supersonic region, such that the factor
$\frac{\Gamma-1}{\Gamma}$ becomes close to unity, and  

\begin{equation}
\dot{M} v_{\infty} = \frac{L}{c} \tau .
\end{equation}
Vink \& Gr\"afener (2012) derived a condition for the wind efficiency number $\eta$:

\begin{equation}
\eta = \frac{\dot{M} v_{\infty}}{L/c} = \tau = 1 .
\label{eq_eta}
\end{equation}
The key point is that one can employ the unique condition $\eta = \tau = 1$ 
right at the transition
from optically thin O-star winds to optically-thick WR winds. 
In other words, if one were to have a data-set containing 
luminosities for 
O and WR stars, the transition mass-loss rate $\dot{M}_{\rm trans}$ is obtained 
by simply considering
the transition luminosity $L_{\rm trans}$ and the terminal velocity $v_{\infty}$
representing the transition point from O to WR stars:

\begin{equation}
\dot{M}_{\rm trans} = \frac{L_{\rm trans}}{v_{\infty} c}
\label{eq_transm}
\end{equation}
This transition point can be obtained by purely spectroscopic means, 
{\em independent} of any assumptions regarding wind clumping.

As $\Gamma = g_{\rm rad}/g$ is expected to be connected to the 
ratio $(v_{\infty} + v_{\rm esc})/v_{\rm esc} = v_\infty/v_{\rm esc} + 1$, and 
$f \simeq \frac{\Gamma - 1}{\Gamma}$, 
Vink \& Gr\"{a}fener followed a model-independent approach, 
adopting $\beta$-type velocity laws, as well as 
full hydrodynamic wind models, computing the 
integral $\tau = \int_{r_s}^{\infty} \kappa \rho\,{\rm d}r$  
numerically using the flux-mean opacity $\kappa_{\rm F}(r)$. 
The mean opacity $\kappa_F$ follows from the resulting
radiative acceleration $g_{\rm rad}$ 

\begin{equation}
g_{\rm rad}(r) = \kappa_F(r)\frac{L}{4\pi c r^2}.
\end{equation}
Whilst $g_{\rm rad}$ follows from the prescribed density $\rho(r)$ and velocity
structures $v(r)$ -- via the equation of motion:

\begin{equation}
  v \frac{{\rm d}v}{{\rm d}r} = g_{\rm rad} - \frac{1}{\rho}\frac{{\rm d}p}{{\rm d}r}
-\frac{GM}{r^2},
\end{equation}
where a grey temperature structure can be 
assumed to compute the gas pressure $p$. 
The sole assumption entering this analysis is that the winds 
  are radiatively driven. 
The resulting mean opacity $\kappa_F$
thus captures all physical effects that could affect
  the radiative driving, including clumping and porosity.
The obtained values for the correction factor  
is 0.6 $\pm$ 0.2. The transition between O and WR
spectral types should in reality occur at:

\begin{equation}
  \dot{M} = f \frac{L_{\rm trans}}{v_{\infty} c} \simeq 0.6 \dot{M}_{\rm trans}.
\end{equation}
There is 
a transition between O and WR spectral types. 
The {\em spectroscopic} transition for spectral
  subtypes O4-6If+ occurs at  $\log(L)=6.05$ and
  $\log(\dot{M}_{\eta=1}/\msunyr)=-4.95$.
This is the transition mass-loss rate for the Arches cluster. 
The only remaining uncertainties are due to 
uncertainties in the terminal velocity and the stellar luminosity $L$, 
with potential errors of at most
$\sim$40\%, and several factors lower 
than the order-of-magnitude uncertainties in mass-loss 
rates resulting form clumping and porosity.  

\subsection{Models close to the Eddington limit}
\label{sec:modelsclosetoedd}

\begin{figure}
\begin{center}
 \includegraphics[width=8cm]{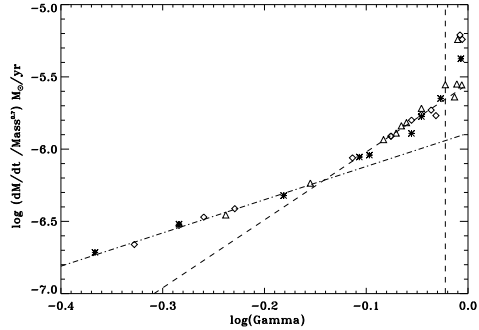} 
\caption{Mass-loss predictions versus the Eddington parameter $\Gamma$ -- divided by $M^{0.7}$. 
Symbols correspond to models of different mass ranges (Vink et al. 2011).}
\label{f_mdotvink}
\end{center}
\end{figure}

The predictions of the O star recipe of Vink et al. (2000) and 
Eq.~(\ref{eq_formula}) are only valid for objects 
at a sufficient distance from the Eddington limit, with $\Gamma$
$\le$ 0.5. There are two regimes where this is no longer the case: (i)
stars that have formed with large initial masses and luminosities, i.e. very
massive stars (VMS) with $M$ $>$ 100\,$M_{\odot}$, and (ii) less extremely
luminous ``normal'' stars that approach the Eddington limit when they have
evolved significantly. Examples of the latter category are LBVs and classical 
WR stars. 

For LBVs, Vink \& de Koter (2002) and Smith et al. (2004) 
showed with Monte Carlo computations that the
mass-loss rate increases more rapidly than Eq.~(\ref{eq_formula}) indicates.
This implies that not only does the mass-loss rate increase when the Eddington limit 
is approached, but the mass-loss rate increases {\it more strongly}, which leads 
to a positive feedback effect on the total mass lost over time.
For VMS, Vink et al. (2011) discovered a kink in the slope of the
mass-loss vs. $\Gamma$ relation at the transition from optically thin O-type
to optically thick WR-type winds. Bestenlehner et al. (2014) performed a homogeneous spectral 
analysis of $>$ 60 Of-Of/WN-WNh stars in 30 Doradus, and confirmed 
the kink empirically. 

Figure\,\ref{f_mdotvink} depicts mass-loss predictions for VMS 
as a function of the Eddington parameter $\Gamma$ from Monte Carlo modelling. 
For ordinary O stars with ``low'' $\Gamma$ the $\dot{M}$ $\propto$ 
$\Gamma^{x}$ relationship is shallow, with $x$ $\simeq$2. 
There is a steepening at higher $\Gamma$, where
$x$ becomes $\simeq$5. 
Here the optical depths and wind efficiencies exceed unity. 

\section{Predictions for low metallicity $Z$ and Pop III stars}

For objects in a $Z$-range representative for the observable Universe 
with $Z/\Zsun$ $>$ 1/100, Monte Carlo mass-loss predictions were provided by 
Vink et al. (2001).
Extending the predictions to extremely low $Z/\Zsun$ $<$
$10^{-2}$, $\dot{M}$ is still expected to drop until the winds 
reach a point where they become susceptible to ion-decoupling and multi-component 
effects (Krticka et al. 2003). In order to maintain a one-fluid wind model is 
by increasing the Eddington factor -- by pumping up the stellar mass
and luminosity. 

For the case of Pop III stars with truly ``zero'' metallicity, i.e. only H and He
present, it seems unlikely that these objects develop
stellar winds of significant strength (Kudritzki et al. 2002; Muijres et al. 2012b).
However, other physical effects may contribute to the driving.  Interesting
possibilities include stellar rotation and pulsations, although pure
vibration models for Pop III stars also indicate little mass loss via
pulsations alone (Baraffe et al. 2001). Perhaps a combination of several effects
could result in large mass loss close to the Eddington limit. Moreover, we
know that even in the present-day Universe a significant amount of mass is
lost in LBV type eruptions, potentially driven by {\it continuum} radiation
pressure, which might be also relevant for the First Stars (Vink \& de Koter 2005; Smith \& Owocki 2006).

Despite the fact that the first generations of massive stars start 
their evolutionary clocks with fewer metals, as the First Stars may be 
highly luminous and/or rapidly rotating, it is not inconceivable that they 
enrich their atmospheres with nitrogen and carbon 
(Meynet et al. 2006), thereby inducing a stellar wind (Vink 2006). 

In a first attempt to investigate the effects of self-enrichment on the
total wind strength, Vink \& de Koter (2005) performed a pilot study of WR mass loss
versus $Z$. The prime interest in WR stars here is that these objects,
especially those of WC subtype, show the products of core burning in their
outer atmospheres. 

\begin{figure}
\begin{center}
   {\includegraphics[width=8cm]{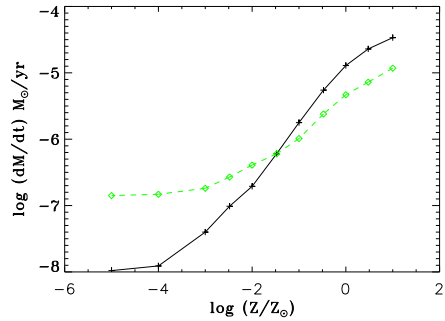}}
\end{center}
\caption{Monte Carlo WR mass-loss predictions as a function of $Z$. The dark
line represents the late-type WN stars, whilst the lighter dashed line
shows the results for late-type WC stars. The slope for the WN models is
similar to the predictions for OB-supergiants, whereas the slope is shallower
for WC stars. At low $Z$, the slope becomes smaller, flattening off
entirely at $Z/\Zsun$ $=$ $10^{-3}$. The computations are from
Vink \& de Koter (2005).} 
\label{wcwn}
\end{figure}

The reasoning behind the assertion that WR winds may not be $Z$-dependent
was that WR stars enrich themselves by burning He into C, and it could
be the large C-abundance that is the most relevant ion for the WC wind
driving, rather than the sheer number of Fe lines. Figure~\ref{wcwn} shows
that despite the fact that the C ions overwhelm the amount of Fe, both
late-type WN (dark line) and WC (light line) show a strong $\dot{M}$-$Z$
dependence, basically because Fe has such a complex electronic structure.

The implications of Fig.~\ref{wcwn} are two-fold. 
First, WR mass-loss rates decrease steeply with $Z$. 
This may be of key relevance for black hole formation and
the progenitor evolution of long duration GRBs. The collapsar
model of MacFadyen \& Woosley (1999) requires a rapidly rotating stellar core prior to 
collapse, but at solar metallicity stellar winds are expected to remove the bulk of 
the core angular momentum (Zahn 1992). The WR $\dot{M}$-$Z$ dependence from
Fig.~\ref{wcwn} provides a route to maintain rapid
rotation, as the winds are weaker at lower $Z$ prior to final collapse.

The second point is that mass
loss is no longer expected to decrease when $Z/\Zsun$ falls 
below $\sim$$10^{-3}$ (due to
the dominance of driving by carbon lines). This 
suggests that once
massive stars enrich their outer atmospheres, radiation-driven 
winds might still exist, even if stars started their lives 
with extremely small amounts of metals.
  
Whether the mass-loss rates are sufficiently high to alter the
evolutionary tracks of the First Stars remains to be seen, but it 
is important to keep in mind that the mass-loss physics does not 
only quasi-linearly depend on $Z$, but that other 
factors, such as the proximity to the $\Gamma$ limit, should also be considered.

\section{Luminous Blue Variables}
\label{sec:lbvs} 

\subsection{What is an LBV?} 

Luminous Blue Variables represent a
short-lived ($\sim 10^{4}-10^{5}$ years) phase of massive star evolution during 
which the objects are subjected to humongous changes in their stellar radii by 
about an order of magnitude. 
They come in two flavors. The largest population of $\sim$30 LBVs in the
Galaxy and the Magellanic Clouds is that of the S\,Doradus variables with
magnitude changes of 1-2 magnitudes on timescales of years to decades (Humphreys \& Davidson 1994).
These are the characteristic S\,Dor variations, represented by the
dotted horizontal lines in Fig.~\ref{hrd}. 
The general understanding is
that the S\,Dor cycles occur at approximately constant bolometric luminosity
(which has yet to be proven) -- principally representing temperature
variations. 
The second type of LBV instability involves objects that show
truly giant eruptions with magnitude changes of order $3-5$ during which the
bolometric luminosity most certainly increases. In the Milky Way it is only
the cases of P~Cygni and Eta Carina which have been noted to exhibit 
such giant outbursts.

\begin{figure}
\begin{center}
   {\includegraphics[width=8cm]{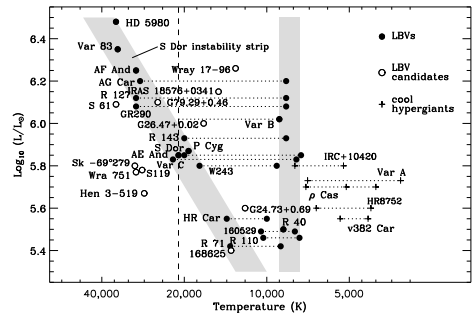}}
\end{center}
\caption{The LBVs in the Hertzsprung-Russell diagram.
The slanted band running from 30~kK at
high $L/\lsun$ to 15~kK at lower luminosity is the
S\,Dor instability strip. The vertical band at a temperature of $\sim$
$8\,000$ K represents the position of the LBVs ``in outburst''. The vertical
line at 21\,000 K is the position of the observed bi-stability jump (Lamers et al. 1995). 
Adapted from Vink (2012) and 
Smith et al. (2004).} 
\label{hrd}
\end{figure}

Whether these types of variability occur in similar or distinct objects
is not yet clear, but in view of the ``unifying'' properties of the 
object P~Cygni it is rather probable that the S\,Dor variables and giant eruptors
are subject to the same type of instabilities near the Eddington limit (see
Vink 2012). 

\begin{figure}
\begin{center}
   {\includegraphics[width=8cm]{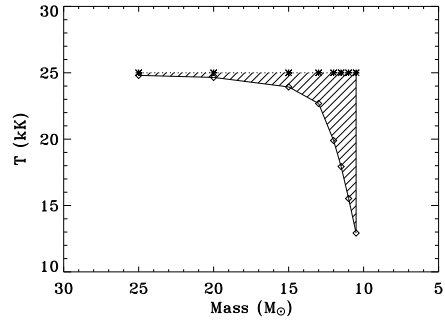}}
\end{center}
\caption{Pseudo-photosphere formation in a relatively low mass
LBV with a high $L/M$ ratio. The difference in inner (dashed) and apparent temperature
(representative for the size of the computed pseudo-photosphere) is plotted
against the stellar mass. These computations have been performed for a
constant luminosity of log $L/\lsun$ $=$ $5.7$. The mass is gradually
decreased whilst the LBV approaches the Eddington limit: the apparent
temperature drops as a result of the lower effective gravity, and the higher 
mass loss results in the formation of a pseudo-photosphere (Smith et al. 2004).} 
\label{app}
\end{figure}

\subsection{Do LBVs form pseudo-photospheres?~~} 

Although it {\it appears} that the photospheric temperatures of the objects in 
Fig.~\ref{hrd} change during HRD transits, there is an alternative possibility
that the underlying star does {\it not}
change its actual temperature but that the star undergoes changes in mass-loss properties 
instead. The second case is normally referred to as the formation of a 
``pseudo-photosphere'' resulting from the formation of an optically thick wind. 

Eta Car's outstanding wind density with \mdot$\sim 10^{-3}$\msunyr 
(Hillier et al. 2001) places $R(\taur=2/3)$ at 80\% of
the terminal velocity, impeding any derivation of the hydrostatic radius, but 
it is not yet clear whether the general LBV population of 
S\,Dor variables have $\dot{M}$ values high enough to produce 
pseudo photospheres. 

As a result of enhanced mass loss during maximum 
it is hypothetically possible to form a pseudo-photosphere. Until the late 1980s this was the leading idea to
explain the colour changes of S\,Dor variables. Using more advanced non-LTE
model atmosphere codes, Leitherer et al. (1989) and de Koter et al. (1996) 
predicted colours based on {\it empirical} LBV mass-loss rates that are not
red enough to make an LBV appear cooler than the temperature of its
underlying surface.  
Despite the proximity of LBVs to the Eddington limit, current consensus is
that LBV winds are generally not sufficiently optically thick. 

Figure~\ref{app}
shows the potential formation of an optically thick wind for a relatively
low-mass (with high $L/M$) LBV in close proximity to the bi-stability jump (Pauldrach \& Puls 1990;
Vink \& de Koter 2002; Groh \& Vink 2011). 
The size of the temperature difference (dashed vs. solid) is a
proxy for the extent of the pseudo-photosphere. The figure demonstrates that
for masses in the range 15-25\,$M_{\odot}$ and fixed luminosity, the winds remain optically thin, but
when the stellar mass approaches values as low as 10\,$M_{\odot}$,
and the star enters the mass-loss regime near the Eddington limit, the photospheric scale-height blows up, 
which results in the formation of a pseudo-photosphere.  

\subsection{Winds during S Doradus variations.~~} 

Although most S\,Dor variables have been subject to photometric
monitoring, only a few have been analysed in sufficient
detail to understand the driving mechanism of their winds.  
Mass-loss rates are of the order of $10^{-3}$ - $10^{-5}$ \msunyr,
whilst terminal wind velocities are in the range $\sim100-500\,\kms$. Obviously, 
these values vary with $L$ and $M$, but there are indications
that the mass loss varies as a function of $\Teff$ when the S\,Dor variables
transit the upper HRD on timescales of years, providing 
an ideal laboratory for testing the theory of radiation-driven
winds.

The Galactic LBV AG\,Car is one of the best monitored and 
analysed S\,Dor variables. 
Vink \& de Koter (2002) predicted $\dot{M}$ rises in line with
radiation-driven wind models for which the $\dot{M}$ variations 
are attributable to ionisation shifts of Fe. Sophisticated non-LTE spectral
analysis have since confirmed these predictions (Groh et al. 2011; Groh \& Vink 2011). 

It is relevant to mention here that this variable wind concept 
(wind bi-stability; see also Pauldrach \& Puls 1990) has been 
suggested to be responsible for circumstellar density variations inferred from 
modulations in radio light-curves and H$\alpha$ spectra of supernovae (Kotak \& Vink 2006; Trundle et al. 2008). 
However most stellar evolution models would have predicted massive stars
with $M$ $\ge$ 25\,$M_{\odot}$ to explode at the end of the WR phase, rather than
after the LBV phase. The implications could be gigantic, 
impacting our most basic understanding of massive star death in the 
Universe (see Smith's Chapter 8).

\subsection{Super-Eddington winds}

Whilst during ``quiet'' phases, LBVs may lose mass 
via ordinary line-driving, some objects, like Eta Car also seem to be 
subject to phases of more extreme mass loss. 
For instance, the giant eruption of $\eta$ Car with a
cumulative loss of $\sim$10\,$M_{\odot}$ between 1840 and 1860 (Smith et al. 2003)
which resulted in the Homunculus nebula corresponds to $\mdot
\approx$~0.1-0.5~\msunyr, which is a factor of 1000 larger than that 
expected from line-driven wind models for an object of that luminosity. 

Shaviv (1998) and Owocki et al. (2004) studied the theory of
porosity-moderated continuum driving in objects that formally 
exceed the Eddington limit. It is possible that 
continuum-driven winds in super-Eddington stars reach mass-loss rates 
close to the {\it photon tiring limit}, $\mdot_{\rm tir} = \Lstar$/ $(G \Mstar/\Rstar)$, 
which could result in a stagnating flow that may lead to spatial structure (van Marle et al. 2008). 
However, it should be noted that alternatively, wind clumping may be the result of 
other instabilities, possibly related to the presence of the Fe opacity peak (Cantiello et al. 2009; 
Gr\"afener et al. 2012; Gr\"afener \& Vink 2013; Glatzel et al. 1993), especially for objects approaching 
the $\Gamma$-limit. 

The {\it general} equation of motion for 
a stellar wind (ignoring gas pressure) is given by: 

\begin{equation}
v\Bigl(1-\frac{a^2}{v^2}\Bigr)\frac{{\rm d}v}{{\rm d}r} \simeq
g_{\rm grav}(r) + g_{\rm rad}(r) =-\frac{GM}{r^2}(1-\Gamma(r)) .
\end{equation}
At the sonic point, $r_s$: $v=a$, and thus 
$g_{\rm rad} = -g_{\rm grav}$ implying $\Gamma(r_s)=1$.
Thus, $\Gamma(r)$ must be $< 1$ below the sonic point 
and $\Gamma(r)$ must be $> 1$ above the sonic point. 
An accelerating wind solution 
thus implies an increasing opacity$\frac{{\rm d}\bar \kappa}{{\rm d}r}|_s>0$ 
(given that $\Gamma(r) = \frac{\bar \kappa(r) L_\ast}{4\pi GMc}$).

If, on the other hand, the entire 
atmosphere is super-Eddington, i.e. $\Gamma(r) >1$ throughout the atmosphere, 
continuum driving might nonetheless become possible.
The reason is that when atmospheres exceed the Eddington limit, instabilities may 
arise which could make them clumpy: outward travelling photons
may avoid regions of enhanced density, which means that the 
medium may behave in a porous manner, leading to a lower
$g_{\rm rad}$. 
This means that the effective Eddington parameter 
can drop below unity. 
However, further out in the wind, the clumps become 
optically thinner as a result of expansion, and the porosity effect decreases.
$\Gamma_{\rm eff}^{\rm cont}$ can now become larger than unity. 
In other words, a wind 
solution with $\Gamma_{\rm eff}^{\rm cont}$ crossing unity is feasible, even when
the stars are formally above the Eddington limit.

Owocki et al. (2004) expressed the effective opacity in terms 
of the so-called porosity length (see Sect.\,8). They showed 
that $\dot{M}$ might 
become substantial when the porosity length is of the order of the 
pressure scale height $H$. Owocki et al. developed the concept
of a power-law distributed porosity length (in analogy to CAK-type 
the line-strength distribution function), and showed 
that even the gigantic mass-loss rate during Eta Car's 
giant eruption might be explained by some form of radiative driving. 

\section{Observed wind parameters}
\label{sec:obswindpara}

Radiation-driven wind models can provide predictions for 
two global wind parameters: the mass-loss rate, \mdot, and the terminal velocity, \vinf.
Most studies rely on the assumption of a smooth 
wind.  
The mass-loss rate then follows from the continuity equation
(Eq.~\ref{eq_const}), and most diagnostics 
are based on a wind model with a prescribed $\beta$-type velocity field.

A useful concept involves the optical-depth invariant $Q$ parameter
(Puls et al. 1996), where 
$Q_{\rm res}$ can be utilised for resonance lines with line opacity $\propto \rho$. 
Alternatively, recombination is a 2-body process and $Q_{\rm rec}$
is useful for recombination based line processes such as H$_{\alpha}$ 
which thus have opacities $\propto \rho^2$,
\beq Q_{\rm res} = \frac{\mdot}{\Rstar \vinf^2}, \qquad \qquad Q_{\rm rec} =
\frac{\mdot}{(\Rstar \vinf)^{1.5}} .
\label{qq} 
\eeq 
Most diagnostics rely on the use of non-LTE
model atmospheres. Stellar and wind parameters, such as 
$\dot{M}$ can be determined by fitting 
resonance and recombination lines simultaneously.
Smooth wind models constitute the ideal case, but the optical depth
invariant $Q_{\rm rec}$ as defined in Eq.~\ref{qq} can easily 
be modified for the case that the winds are 
clumped (Sect.~\ref{sec:clumping}, Eq.~\ref{qclump}). 

A more detailed discussion of the various methods to
derive wind parameters is given in Puls et al. (2008). 
The most common line profiles in a stellar wind are 
(i) UV P\,Cygni
profiles with a blue absorption trough and a red emission peak, and  
(ii) optical emission lines (such as H$\alpha$). 
These line shapes are caused by different population mechanisms of
the upper energy level of the transition. 
In a P~Cygni scattering line, the upper level is populated by the balancing act 
between
absorption from and spontaneous decay to the lower level. 
An emission line is formed if the upper level is populated by recombinations 
from above (see however Puls et al. 1998; Petrov et al. 2014 for the formation 
of P\,Cygni H$\alpha$ lines in the cooler BA supergiants). 

\subsection{Ultraviolet P\,Cygni Resonance lines}

P\,Cygni lines may be used to
determine the velocity field in stellar winds, and in particular \vinf.
\Ha\ is generally utilised to derive $\dot{M}$ (or $Q_{\rm rec}$).
UV P-Cygni lines from hot stars (e.g. \CIV\ and
\PV) are usually analysed by means of the Sobolev optical 
depth:
\beq
\taus(r)  =  \frac{\frac{\pi e}{m_{\rm e} c} f \nl(r) \lambda}{\dvdr} \frac{\Rstar}{\vinf},
\label{taus}
\eeq
where $f$ is the oscillator-strength and $\nl$ the lower occupation number of the
transition. Relating the occupation number, \nl, to the density:
\beq
\label{kdef}
\taus(r)  = \frac{1}{r^2 v \dvdr}\,E(r)q(r) \, \frac{\mdot}{\Rstar\vinf^2} \,
\frac{(\pi e^2)/(m_{\rm e}c)}{4\pi m_{\rm H}} \,
\frac{A_{\rm k}}{1+4Y} \, f \lambda  ,
\eeq
where $E$ is the excitation factor of the lower level, $q$ the ionisation
fraction, $A_{\rm k}$ the abundance of the element, 
and $Y$ the He abundance. This quantity is 
invariant with respect to $Q_{\rm res} = \mdot /(\Rstar
\vinf^2)$ (see Eq.~\ref{qq}) as long as the ground-state population 
is proportional to the density $\rho$. 
Thus, $\dot{M}$ can be derived from resonance line 
P\,Cygni profiles when the ionisation fraction is known. 
Most P\,Cygni lines however are 
saturated and mass-loss rate derivations become
unfeasible, such that only lower limits on $\dot{M}$ can be determined. 

UV resonance lines have been
considered relatively clean from clumping effects, but 
this might not be the case if porosity effects become important.

\subsection{The H$\alpha$ recombination emission line}

The most oft-used diagnostics 
to derive \mdot\ for O-star winds 
involves \Ha, for which there is hardly any uncertainty due to 
ionisation. 
The \Ha\ opacity scales with $\rho^2$, and
\beq
\taus(r)\propto
\frac{\mdot^2}{(\Rstar \vinf)^3} \frac{b_2(r)}{r^4 v^2 \dvdr},
\eeq
i.e., the scaling invariant quantity is now $Q_{\rm rec}^2$ (Eq.~\ref{qq}), and $b_2$ is the non-LTE
departure coefficient of $n_2$.

The challenge with \Ha\ concerns 
its $\rho^2$ dependence. Any
notable inhomogeneity will necessarily result in an 
\mdot\ overestimate if clumping is neglected in the analysis. An advantage 
is the fact that \Ha\ remains optically thin in the main part of
the emitting wind, such that porosity effects can be
neglected (which is not the case for UV resonance lines). 

\subsection{Radio and (sub)millimetre continuum emission} 

A somewhat different approach to measure mass-loss rates is to utilize 
long wavelength radio and (sub)millimetre continua. 
In fact this approach may lead to the most accurate results, as they are 
model-independent. 
The basic concept is to measure the excess wind flux over that from the stellar 
photosphere.
This excess flux is emitted by free-free and bound-free processes. The reason 
the excess flux
becomes more important at longer (sub)-mm/radio wavelengths is 
due to the $\lambda^2$ dependence
of the opacities. 

Following Wright \& Barlow (1975), Panagia \& Felli (1975),  
and Lamers \& Cassinelli (1999), the dominant free-free 
opacity (in units of cm$^{-1}$) at frequency $\nu$ can be 
written as:
\beq 
\kappa_\nu \propto n_{\rm i} \, n_{\rm e} \, g_{\nu} \, (\frac{1}{\nu^2}) 
            \propto {\frac{\dot{M}}{\vinf}}^2 \, (\frac{1}{r^4}) \,  g_{\nu} \, (\frac{1}{\nu^2}), 
\eeq
in cm$^{-3}$, and $g_{\nu}$ is the Gaunt factor for free-free emission.
For an isothermal wind and frozen-in ionisation,
$\bar{z}$ (the mean value of the atomic charge) and 
$\mu_{\rm e}$ and $\mu_{\rm i}$ remain constant, and:
\beq
\kappa_\nu \propto {g_{\nu} \lambda^2 \rho^2},
\eeq
which increases with $\lambda$ and $\rho$. As  
the continuum becomes 
optically thick in the wind in free-free opacity 
the emitting wind volume increases as a function 
of $\lambda$, leading to the formation of a radio photosphere
where the the radio emission dominates the stellar photospheric emission. 
For a typical
O supergiant this occurs at about 100 stellar radii.
At such large distances the outflow reaches its terminal wind velocity 
and an analytic solution 
of the radiative transfer problem
becomes possible:
\beq
\label{fnuradio}
F_\nu \propto \Bigl(\frac{\mdot}{\vinf}\Bigr)^{4/3}\, 
\frac{\bigl(\nu g_{\nu}\bigr)^{2/3}}{d^2} ,
\eeq
where $F_\nu$ is the observed radio flux measured in Jansky, 
\mdot\ in units of \msunyr, \vinf\ in \kms, distance
$d$ to the star in kpc and frequency $\nu$ in Hz. 
Thus, the spectral index
of {\it thermal} wind emission is close to 0.6.

\section{Wind clumping}
\label{sec:clumping}

H$\alpha$ and long-wavelength continuum 
diagnostics depend on the density squared, and are thus
sensitive to clumping, whereas UV P Cygni lines such as 
P{\sc v} are insensitive to clumping, as they depend linearly 
on density. In the canonical optically thin (micro-clumping) approach 
the wind is divided into a portion of the wind that 
contains all the material with a volume filling factor $f$
(the reciprocal of the clumping factor $D$), whilst the remainder of the
wind is assumed to be void. 
In reality however, clumped winds are porous with a range of clump
sizes, masses, and optical depths. 

Wind clumping has been extensively discussed for canonical 
20-60\,\msun\ O-type stars and 
WR stars in a dedicated clumping workshop (Hamann et al. 2008). Here
one may also find studies of X-ray observations (see also 
Cohen et al. 2014 and references therein for more recent work). 

\subsection{Optically thin clumping (``micro-clumping'')}
\label{sec:clumpobs}

The general concept of optically-thin micro clumping is simply 
based on the assumption that the wind is made up of large 
numbers of small-scale density
clumps. Largely motivated by the results from
hydrodynamic simulations including the line-deshadowing 
instability (LDI; see Owocki's Chapter 5), the
inter-clump gas is usually assumed to be void. 
The average density $\rhob = \mdot/(4 \pi r^2 v$) is given by:
\beq 
\rhob = f \rho_{C},\quad \rhobtwo = f (\rho_{C})^2 
\eeq 
where $\rho_{C}$ is the density inside the over-dense clumps, and \rhobtwo\ is
the mean of the squared density. Thus, the clumping factor: 
\beq 
D
= \rhobtwo/\rhob^2 \quad \Rightarrow \quad D=f^{-1} \quad \mbox{and}
\quad \rho_{C} = D \rhob, 
\eeq 
measures the clump over-density. As the inter-clump space is assumed to be void, 
matter is only present inside the clumps, with density $\rho_{C}$, and with its 
opacity given by $\kappa=\kappa_C(D \rhob)$, where $C$
represents the quantities inside the clump. Optical depths may be 
calculated via $\tau=\int \kappa_C(D \rhob) f \dd r$ with a reduced
path length $(f \dd r)$ as to correct for the volume where  
clumps are actually present. 

The formulation is only correct as long as the clumps are optically thin, and 
optical depths may be expressed by a mean opacity $\bar \kappa$:
\beq 
\bar \kappa =
\kappa_C(D \rhob)f = \frac{1}{D} \kappa_C(D \rhob).  
\label{kappabar}
\eeq 
Thus, for processes that are linearly dependent on density, the mean 
opacity of a clumped medium is exactly the same as for a smooth wind, 
whilst for processes that scale with the density squared, 
mean opacities are enhanced by the clumping factor D.

It should be noted that processes 
described by the optically thin micro-clumping approach do not
depend on clump size nor geometry, but only the sheer 
clumping {\it factor}. The enhanced opacity
for $\rho^2$ dependent processes implies that $\dot{M}$ 
derived by such diagnostics are a factor of $\sqrt{D}$ lower than
older mass-loss rates derived with the assumption of smooth winds. 
As a result, the optical depth invariant, $Q_{\rm rec}$ (see Eq.~\ref{qq}) 
transforms into:
\beq
Q_{\rm rec} = \frac{\mdot \sqrt{D}}{(\Rstar \vinf)^{1.5}}.
\label{qclump}
\eeq
Note that also for the case of thermal radio and (sub)-mm
continuum emission the scaling invariant is proportional to 
$\mdot/\Rstar^{1.5}$, i.e. 
very similar to $Q_{\rm rec}$ for optical emission lines, such as
H$\alpha$.
Abbott et al. (1981) studied the effects of clumping on
the wind radio emission as a function of
the volume filling factor and the density ratio between
clumped and inter-clump material. For the standard assumption of 
vanishing inter-clump density, Abbott et al. showed that the radio flux may be
a factor $f^{-2/3}$ larger than that from a smooth wind with the same
$\dot{M}$. In other words, using Eq.~\ref{fnuradio}, it can be noted that 
radio mass-loss rates derived from clumped winds must also be lower 
than those derived from smooth winds.

\subsection{The \PV\ problem} 

Due to the very low cosmic abundance of phosphorus (P), the
\PV\ doublet remains unsaturated, even when P$^{+4}$ is
dominant. 
This allows for a direct estimate of the product
\mdot\qb, where \qb\ is a spatial average of the ion fraction. 
Unfortunately, \qb\ estimates for a given
resonance line are uncertain due to shocks and associated  
X-ray ionisation.  
Empirical determination of ionisation fractions is normally 
not feasible, as resonance lines from consecutive ionisation stages 
are not generally available.
Nevertheless, for \PV, insight is gained from FUSE data:
for those O-stars in a certain 
$\qb \simeq 1$ region, the \PV\ line should provide an accurate 
estimate of \mdot, as the pure linear character 
with $\rho$ makes it clumping independent.

Fullerton et al. (2006) selected a large sample of O-stars, which also had 
$\rho^2$ (from H$\alpha$/radio) estimates available,
and compared both $\rho$-linear UV and $\rho$-quadratic dependent 
methods. They found enormous discrepancies, with a median 
\mdot($\rho^2$)/(\mdot(\PV)\qb) = 20 in mid-O supergiants, implying 
an extreme clumping factor D $\simeq$ 400 {\it if} the winds could indeed 
be treated in an optically thin (micro-clumping) approach (see also Bouret et al. 2003). 

\begin{figure}
\begin{center}
   {\includegraphics[width=8cm]{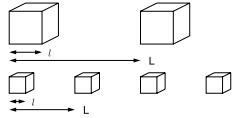}}
\end{center}
\caption{Schematic explanation of porosity, involving a notable 
difference between the volume filling fraction f (and its reciprocal clumping factor $D = 1/f$), 
which is the same for the top and bottom case, and the separation of the clumps $L$, which is 
larger in the top case than the bottom case (from Muijres et al. 2011).}
\label{por}
\end{figure}

\subsection{Optically thick clumping (``macro''-clumping)} 
\label{sec:clumptreat}

With studies yielding clumping factors ranging from $D$ up to 400, one 
may wonder whether a pure micro-clumping analysis is physically sound. 
Most of the atmospheric codes only consider density
variations, but hydrodynamic
simulations also reveal strong velocity changes inside the
clumps. 
Most worrisome is probably the assumption 
that all clumps are assumed to be optically thin. 

Within the optically thin approach, a clump 
has a size smaller than the photon mean free path. 
However, in an optically thick clump, photons may interact with the gas 
several times before they escape through the inter-clump gas. 
Whether a
clump is optically thin or thick depends on the abundance, ionisation
fraction, and cross-section of the transition. 

For optically thick clumps, photons care about the distribution,
the size and the geometry of the clumps (see Fig.\,\ref{por}). 
The conventional description of macro-clumping is based on a  
clump size, $l(r)$, and an average spacing of a statistical
distribution of clumps, $L(r)$, which are related to $f$:
\beq
f = \bigl(\frac{l}{L}\bigr)^3 = \frac{1}{D}.
\label{eq:fmacroclu}
\eeq
Following Eq.~\ref{kappabar}, the optical depth across
a clump of size $l$ and opacity $\kappa_C$ becomes:
\beq
\tau_C = \kappa_C l = \bar \kappa D l = \bar
\kappa \frac{L^3}{l^2} = \bar \kappa h,
\label{eq:taumacroclu}
\eeq
with mean opacity $\bar \kappa$ (Eq.~\ref{kappabar}) and porosity length
$h=L^3/l^2$. The
porosity length $h$ involves the key parameter to define 
a clumped medium, as $h$ corresponds to the photon mean free path in a
medium consisting of optically thick clumps.

The effective clump cross section, i.e., the {\it spatial} 
cross section now corrected for the fraction of 
transmitted radiation, becomes:
\beq
\sigma_C = l^2 \, (1 - \eu^{-\tau_C}),
\label{eq:croscluthick}
\eeq
and the effective opacity becomes:
\beq
\kappa_{\rm eff} = n_C \sigma_C = \frac{l^2\,(1 - \eu^{-\tau_C})}{L^3} = 
\bar \kappa \frac{(1 -
\eu^{-\tau_C})}{\tau_C},
\label{eq:chicluthick}
\eeq
where $n_C$ is the clump number density. 
The key point is that this very equation
holds for clumps of any optical thickness! 
For instance, in the optically thin limit, the micro-clumping approximation 
is recovered: $\kappa_{\rm eff} = \bar \kappa$, which depends on $f$ and not 
on clump size or distribution. In the optically
thick case, the effective opacity is indeed reduced appropriately, 
$\kappa_{\rm eff} = \bar \kappa/\tau_C$ = $h^{-1}$ now only depending on $h$.

\begin{figure}
\begin{center}
\centerline{ \hspace{-0.2cm}
\includegraphics[width=11.5cm,angle=0]{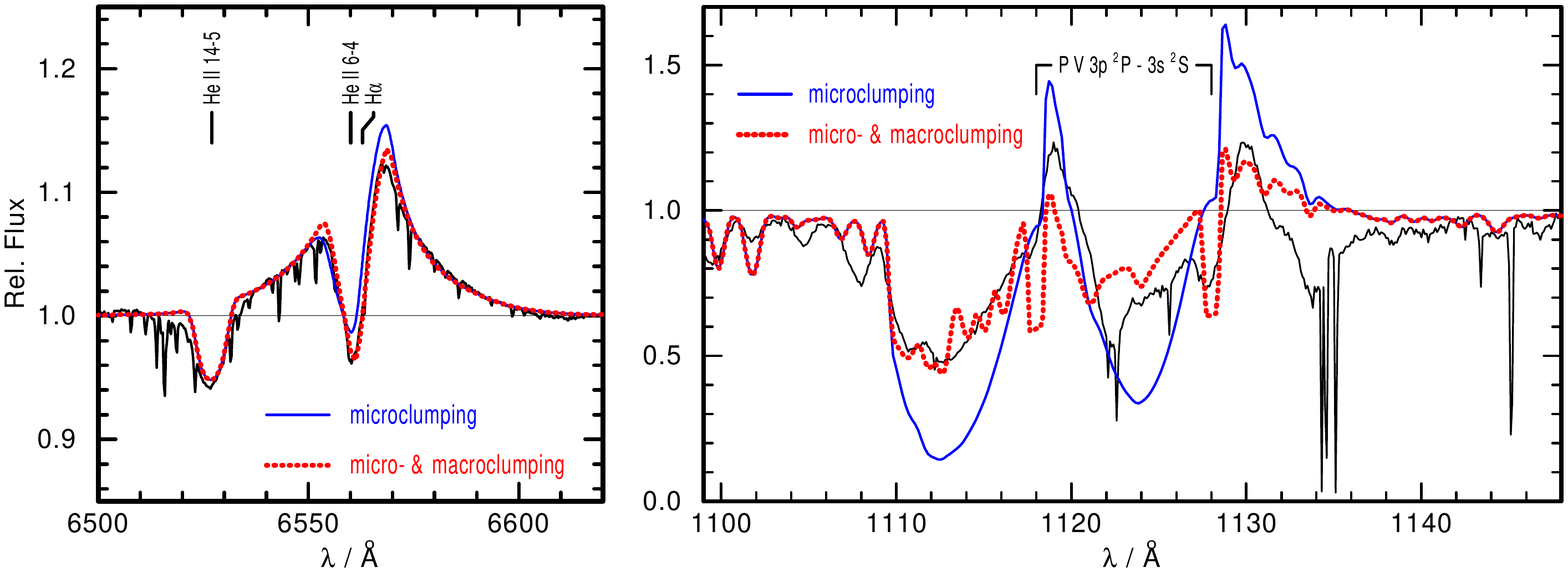}}
\vspace{-0.1cm} 
\caption{Porosity as a possible solution for the PV problem.
Adapted from Oskinova et al. (2007).} 
\label{fig:oskiclump}
\end{center}
\end{figure}

Oskinova et al. (2007) 
employed the effective opacity concept 
in the formal integral for the  
line profile modelling of the O supergiant $\zeta$~Pup.
Figure~\ref{fig:oskiclump} shows that the most pronounced 
effect involves strong resonance lines, such as \PV\ which can be
reproduced by this macro-clumping approach -- without the need for extremely 
low $\dot{M}$ -- resulting from an effective opacity reduction 
when clumps become optically thick. Given that \Ha\ remains 
optically thin for O stars it is not affected by porosity\footnote{This might be different for B supergiants below 
the bi-stability jump (see Petrov et al. 2014).}, and 
it can be reproduced simultaneously with \PV. This enables 
a solution to the \PV\ problem (see also Surlan et al. 2013). 

However, this porosity concept was developed 
for continuum processes, whilst line processes may also 
be affected by velocity-field changes.
Owocki (2008) performed LDI simulations 
where the line strength was described through a velocity-clumping factor. 
These simulations resulted in a reduced wind absorption due to 
porosity in velocity space, which has been termed ``vorosity''.  
The issue with explaining a reduced \PV\ line-strength
through vorosity is that one needs to have a relatively 
large number of substantial velocity gaps, which does not 
easily arise from the LDI simulations. 
In any case, there is still a need to study
scenarios including both porosity and vorosity, as well as 
how they interrelate (Sundqvist et al. 2012).

\subsection{Quantifying the number of clumps}
\label{tclump}

\
\begin{figure}
\begin{center}
   {\includegraphics[width=8cm]{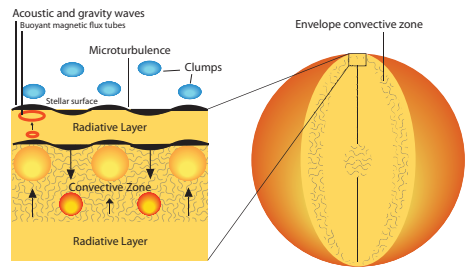}}
\end{center}
\caption{Cartoon of the physical processes involved in sub-surface convection. 
Acoustic and gravity waves are emitted in the convective zone, and 
travel through the radiative layers, reaching the stellar surface, thereby 
inducing density and velocity fluctuations. 
In this picture, clumping starts at the wind base. 
From Cantiello et al. (2009).}
  \label{sketch}
\end{figure}

\noindent
In the traditional view of line-driven winds of O-type stars via the 
CAK theory and the associated LDI, clumping would be expected to develop
in the wind when the wind velocities are large enough to produce shocked
structures. For typical O star winds, this is thought to occur at 
about half the terminal wind velocity at about 1.5 stellar radii.

Various observational indications, involving the existence of linear 
polarisation (e.g. Davies et al. 2005) as well as radial dependent 
spectral diagnostics (Puls et al. 2006) however show that clumping must 
already exist at very low wind velocities, 
and more likely arise in the stellar photosphere.
Cantiello et al. (2009) suggested that waves produced by the subsurface 
convection zone associated with the Fe opacity peak could 
lead to velocity fluctuations, and possibly density fluctuations, and 
thus be the root cause for the observed wind clumping at the
stellar surface (see Fig.\,\ref{sketch}). 

Assuming the horizontal extent of the clumps to be comparable to the
sub-photospheric pressure scale height $H_{\rm p}$, one may
estimate the number of convective cells by dividing the stellar surface area
by the surface area of a convective cell finding that it scales as
($R/H_{\rm P})^2$. For main-sequence O stars in the canonical
mass range 20-60\,$M_{\odot}$, pressure scale heights are within the range
0.04-0.24 $R_{\odot}$, corresponding to a total number of clumps
6 $\times 10^3-6 \times 10^4$. These estimates
may in principle be tested through
linear polarisation variability, which probes wind asphericity
at the wind base.

In an investigation of WR linear polarisation variability 
Robert et al. (1989) uncovered an 
anti-correlation between the wind terminal 
velocity and the scatter in polarisation. They  
interpreted this as the result of blobs that grow or survive 
more effectively in slow winds than fast winds. 
Davies et al. (2005) found this trend to
continue into the regime of LBVs, with even lower $\vinf$. LBVs are 
are thus an ideal test-bed for constraining clump properties, due
to the larger wind-flow times.
Davies et al. showed that over 50\% of LBVs are intrinsically polarised. 
As the polarisation angle was found to vary 
irregularly with time, the polarisation line effects were attributed to 
wind clumping. 
Monte Carlo models for scattering off wind clumps have 
been developed by Code \& Whitney (1995); Rodriguez \& Magalhaes (2000); and Harries (2000), 
whilst analytic models to produce the variability of the linear polarisation 
may be found in Davies et al. (2007); Li et al. (2009); and Townsend \& Mast (2011). 

An example of an analytic model that predicts the time-averaged
polarisation for the LBV P\,Cygni is presented in 
Fig.~\ref{fig:pcyg_thick}. The clump ejection rate per wind flow-time
$\mathcal{N}$ is defined as ${\mathcal N} = \dot{N} t_{\rm fl} = \dot{N}
R_{\star}/v_{\infty}$, where the clump ejection rate, $\dot{N}$, is related
to $\dot{M}$ as $\dot{M} = \dot{N} N_{e} \mu_{e} m_{H}$, where
$N_{e}$ is the number of electrons in each clump, and $\mu_{e}$ is 
the mean mass
per electron. There are two
regimes where the observed polarisation level can be achieved.
One is where the ejection rate is low and a few very optically thick
clumps are expelled; the other one is that of a very large number of clumps. 
These two cases can be
distinguished via time resolved polarimetry. Given the relatively short
timescale of the observed polarisation variability, Davies et al. argued 
that LBV winds consist of order 
thousands of clumps near the photosphere. 

Nevertheless, for main-sequence O stars the derivation of 
wind-clump sizes from polarimetry 
has not yet been feasible as very high signal-to-noise data 
are required. 
LBVs however provide 
an excellent group of test-objects owing to the 
 combination of higher mass-loss rates, and lower terminal wind velocities. 
Davies et al. (2007) showed that in order to produce the observed 
polarisation variability of P\,Cygni, the wind 
should consist of $\sim$ 1000 clumps per wind flow-time. 
In order to check whether this is compatible with the 
sub-surface convection scenario ultimately being the root cause for 
wind clumping, one would need to consider the sub-surface convective regions 
of an object with global 
properties similar to those of P\,Cygni. Due to the lower LBV gravity, 
the pressure scale height is about 4$R_{\odot}$, i.e. significantly 
larger than for O-type stars. 
As a result, the same estimate for the number of clumps drops to 
about 500 clumps per wind-flow time, which appears to be consistent  
with that derived for P~Cygni from observations (see Fig.\,\ref{fig_pcyg}).

\begin{figure}
\begin{center}
   {\includegraphics[width=8cm]{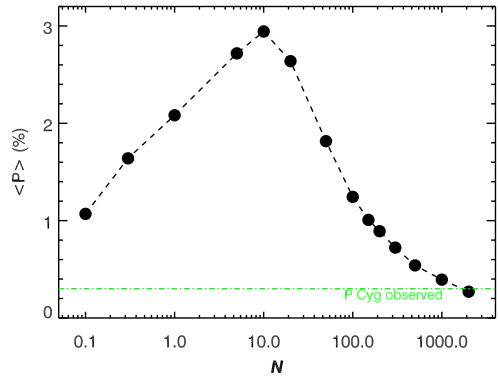}}
\label{fig_pcyg}
\end{center}
\caption{Time-averaged polarisation over a range of ejection rates per
wind flow-time. At $\mathcal{N} \sim 20$, the optical depth
per clump exceeds unity and the overall polarisation falls off (see
Davies et al. 2007 for details). The observed polarisation level for
the LBV P~Cygni is given by the dash-dotted line. There are two
ejection-rate regimes where the required polarisation level can be
achieved.}
\label{fig:pcyg_thick}
\end{figure}

\subsection{Effects on mass-loss predictions}

Muijres et al. (2011) studied the possible effects of
both optically thin and thick wind clumping (porosity)
on mass-loss predictions for O-type stars.

\begin{figure}
\begin{center}
   {\includegraphics[width=8cm]{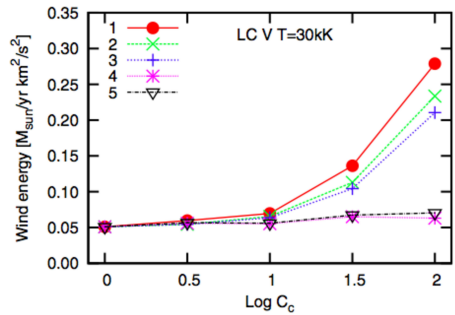}}
\end{center}
\caption{The effect of optically thin micro-clumping on the wind kinetic energy in Monte Carlo simulations
for different clumping stratifications for 30,000 K OV-type stars. 
The smooth wind models have $D = 1$. The numbers 1-5 refer to different clumping stratifications (see Muijres et al. 2011 for details), but 
clumping in the outer winds (stratifications 1 through 3) results in an increase of the kinetic wind energy due to a larger 
number of effective driving lines.}
\label{muij_cl}
\end{figure}

Because of the non-linear character of the equation of motion, the CAK solution is complex, with the
physics
involving instabilities due to the LDI (e.g. Owocki et al. 1988). One of the key implications of
the LDI
is that in hydro-dynamical simulations the time-averaged $\dot{M}$ is {\it not}
anticipated to be affected by wind clumping, as
it has the same average $\dot{M}$
as the smooth CAK solution. However, the shocked velocity structure and its
associated
density structure are expected to result in effects on the mass-loss
diagnostics. 

In contrast to the LDI simulations,
Muijres et al. (2011) studied the effects of clumping
on $g_{\rm rad}$ due to changes
of the ionisation structure, as well as the effects of wind porosity, using
Monte Carlo simulations. 
When only accounting for optically thin (micro) clumping $g_{\rm rad}$ was found to {\it increase} for
certain clumping stratifications $D(r)$, but only
for an extremely high clumping factor of $D\sim100$ (see Fig.\,\ref{muij_cl} for a range of 
clumping factors and stratifications).
The reason $g_{\rm rad}$ may increase is the result of 
recombination yielding more flux-weighted opacity
from lower Fe ionisation stages (similar to the bi-stability physics).
For $D=10$ the effects were however found to be relatively minor.

When simultaneously also accounting for optically thick
(macro) clumping, the effects were partially reversed, as photons could 
now escape in between the clumps without interaction, and the predicted
$g_{\rm rad}$ goes down, as well as up (see Fig.\,\ref{muij_por} for a range of 
clumping stratifications). Nevertheless, again,
for $D =10$ the effects were found to be rather modest.

\begin{figure}
\begin{center}
   {\includegraphics[width=8cm]{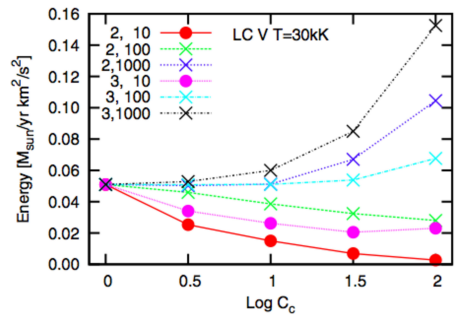}}
\end{center}
\caption{The effects of optically thick macro-clumping on the wind kinetic energy in Monte Carlo simulations
for different clumping {\it and porosity} stratifications for 30,000 K OV-type stars.
The smooth wind models have $D = 1$. 
The numbers 1-5 refer to different clumping stratifications (see Muijres et al. 2011 for details).}
\label{muij_por}
\end{figure}

A fully consistent study of the impact of
wind-clumping on predicted wind properties has yet to be performed. 

\section{Summary}

As we mentioned in Sect.\,1 (see also Chapters 6 and 7) the evolution and fate 
of VMS are predominantly determined by $\dot{M}$.
Current stellar evolution models for VMS (e.g. Yusof et al. 2013; 
K\"ohler et al. 2014) utilise the smooth 
Monte Carlo {\it theoretical} predictions of Vink et al. (2000).

However, it has become clear that {\it empirical}
$\dot{M}$ rates have been overestimated when determined from 
$\rho^2$ diagnostics such as \Ha. 
According to Repolust et al. (2004) and Mokiem et al. (2007) the non-clumping 
corrected empirical rates are a factor 2-3 {\it higher} 
than the Vink et al. (2000) rates, meaning that 
moderate clumping effects (with $D$ = 4-10) are 
indirectly accounted for in stellar evolution models, noting that 
all recently reviewed stellar models 
employ Vink et al. (2000) rates according to Martins \& Palacios (2013). 
 
However, there has been a breakthrough in our understanding 
of $\dot{M}$ for VMS in close proximity to the Eddington $\Gamma$ limit. 
Vink et al. (2011) discovered a ``kink'' in the $\dot{M}$ 
vs. $\Gamma$ relation at the transition from optically thin O-type
to optically thick winds.
For ordinary O stars with ``low'' $\Gamma$ the $\dot{M}$ $\propto$
$\Gamma^{x}$ relationship is shallow, with $x$ $\simeq$2.
There is a steepening at higher $\Gamma$, where
$x$ becomes $\simeq$5. This mass-loss enhancement due to VMS
in proximity to the $\Gamma$-limit has not yet been included 
in evolutionary models of VMS, and is likely to be crucial
for their ultimate fate.

We also discussed a methodology that involves a model-{\it independent} $\dot{M}$ 
indicator: the transition mass-loss 
rate $\dot{M}_{\rm trans}$ -- located 
right at the transition from optically thin to optically thick stellar winds (Vink \& Gr\"afener 2012). 
As $\dot{M}_{\rm trans}$ is model independent, {\it all} that is required 
is to postulate
the {\it spectroscopic} transition point in a given data-set and to determine 
the far more accurate $L$ parameter.
In other words $\dot{M}_{\rm trans}$ is extremely useful for 
calibrating wind mass loss, and assessing its role in
mass loss during stellar evolution.
As was mentioned, current stellar models use Vink et al. 
mass-loss rates that have been reduced by factors of 2-3 compared
to previous unclumped empirical rates, and there is thus no immediate 
reason to reduce them further, unless clumping factors 
would be higher than $\sim$10.

Furthermore, we have also seen in Sect.\,8 that clumping can affect 
the Monte Carlo mass-loss predictions in various ways, involving
both reductions and increases in $\dot{M}$. We have also  
highlighted that both the origin and onset of wind clumping remain unclear. 
Polarisation measurements call for clumping to be already present 
in the stellar photosphere, but how this would interact with 
the hydro-dynamical LDI simulations further out, and how this would 
need to be consistently incorporated into 
radiative transfer calculations and mass-loss predictions is as yet 
unclear. For these reasons, the search for the 
nature and implications of wind clumping should continue!

\end{document}